\documentclass[acmsmall]{acmart}

\AtBeginDocument{%
  \providecommand\BibTeX{{%
    \normalfont B\kern-0.5em{\scshape i\kern-0.25em b}\kern-0.8em\TeX}}}

\setcopyright{acmlicensed}
\acmJournal{PACMHCI}
\acmYear{2022} \acmVolume{6} \acmNumber{CSCW1} \acmArticle{74} \acmMonth{4} \acmPrice{15.00}\acmDOI{10.1145/3512921}

\usepackage[normalem]{ulem}
\usepackage{xcolor}
\usepackage{amsthm}
\usepackage{siunitx}
\usepackage[inline]{enumitem}
\usepackage{makecell}
\usepackage{multirow}
\usepackage{longtable}
\usepackage{subcaption}
\usepackage{makecell}
\usepackage{titlesec}

\usepackage[super]{nth}

\newcommand{\rev}[1]{#1}
\newcommand{\camready}[1]{#1}

\newtheorem*{resques}{Research Question}
\newtheorem{hyp}{Hypothesis}
\begin{document}

\title{A Web-Scale Analysis of the Community Origins of Image Memes}

\author{Durim Morina}
\orcid{0000-0002-9101-8382}
\affiliation{%
  \institution{Stanford University}
  \city{Stanford}
  \country{USA}}
\email{dmorina@cs.stanford.edu}

\author{Michael S. Bernstein}
\orcid{0000-0001-8020-9434}
\affiliation{%
  \institution{Stanford University}
  \city{Stanford}
  \country{USA}}
\email{msb@cs.stanford.edu}

\renewcommand{\shortauthors}{Durim Morina \& Michael S. Bernstein}

\begin{abstract}
    Where do the most popular online cultural artifacts such as image memes originate?
    Media narratives suggest that cultural innovations often originate in peripheral communities and then diffuse to the mainstream core; behavioral science suggests that intermediate network positions that bridge between the periphery and the core are especially likely to originate many influential cultural innovations.
    Research has yet to fully adjudicate between these predictions because prior work focuses on individual platforms such as Twitter; however, any single platform is only a small, incomplete part of the larger online cultural ecosystem.
    In this paper, we perform the first analysis of the origins and diffusion of image memes at web scale, via a one-month crawl of all indexible online communities that principally share meme images with English text overlays.
    Our results suggest that communities at the core of the network originate the most highly diffused image memes: the top $10\%$ of communities by network centrality originate the memes that generate $62\%$ of the image meme diffusion events on the web.
    A zero-inflated negative binomial regression confirms that memes from core communities are more likely to diffuse than those from peripheral communities even when controlling for community size and activity level.
    However, a replication analysis that follows the traditional approach of testing the same question only within a single large community, Reddit, finds the regression coefficients reversed---underscoring the importance of engaging in web-scale, cross-community analyses.
    The ecosystem-level viewpoint of this work positions the web as a highly centralized generator of cultural artifacts such as image memes.
\end{abstract}



\maketitle

\section[intro]{Introduction}\label{sec:introduction}
\rev{Which online communities originated the cultural artifacts that we see around us online? Cultural artifacts are passed from community to community, upvoted and downvoted, remixed and reposted. When we come across an artifact such as an image meme posted to an online community, we might wonder: where did this cultural artifact originate? Which communities are generating the internet culture that we see all around us?}

Common narratives, drawing on examples such as jazz arising in African-American areas of New Orleans~\cite{lopes2002rise} and
the emo fashion movement arising from the post-punk subculture in Washington, D.C.\rev{~\cite{greenwald2003nothing}}, suggest that cultural innovation arises
in fringe, \textit{peripheral} communities that are near the edge of the network.
These innovations are then adapted and popularized by the \textit{core}, or mainstream center of the network.
These narratives recur in online communities as well, with the subversive online community 4chan \slash{}b\slash{} generating some of the most recognizable \camready{early} internet memes such as \mbox{LOLcats} and rickrolling~\cite{bernstein20114chan,keks}.
Organizational sociologists refine this notion, suggesting that cultural innovation often \rev{originates} not in the most peripheral communities, but from those in an intermediate network location between core and periphery, because they are able to draw ideas from the periphery and bridge them into the core~\cite{cattani2008core,doi:10.1287/orsc.1110.0673}. 

However, research still lacks a firm answer as to whether these theories find substantial empirical support online, because research to date has focused on diffusion within individual communities, granting a valuable glimpse but an incomplete picture.
Work focusing on individual communities (platforms, domains, or socio-technical systems) such as Twitter~\cite{Huang:2010:CTT:1810617.1810647,petrovic2011rt}, Digg~\cite{szabo2010predicting}, Facebook~\cite{cheng2014can}, and LiveJournal~\cite{backstrom2006group} provide lenses on how memes are created within that specific community.
However, societies cluster into groups that inhabit different spaces~\cite{mcpherson2001birds}, so any given system is only a small part of the larger ecosystem; we are left with an incomplete picture that leaves out many peripheral communities.
How might these dynamics play out across the broader cultural landscape of the entire web?
Studies of diffusion that cross multiple communities are rare~\cite{zannettou2018origins}, and are typically focused either on politics~\cite{leskovec2009meme,adamic2005political} or on cascade size~\cite{goel2012structure}.
As a result, the question of where image memes and other online cultural artifacts originate is still unanswered.

In this paper, we contribute the first \textit{web-scale} study of image memes: an analysis of the origination and diffusion of image memes that contain English-language text across essentially all indexable communities on the web.
By pursuing a web-scale analysis, we achieve an ecosystem-level viewpoint that affords us a novel opportunity to track image memes as they arise and spread across the web, not just within single communities or across a small number of communities.
\rev{We focus on image memes---static pictures with (often humorous) overlaid text---as a case study of cultural artficts, because image memes are among the dominant cultural icons on the modern web.}
%

Our approach produces a capture of new memes posted to the public, English-speaking web over one month. 
We proceed in two stages: (1)~a \textit{crawl} to identify all public, English-speaking communities on the web that post majority meme content, and (2)~a \textit{high frequency scrape} of those communities over a period of one month to identify new memes and track their diffusion.
The first stage, the crawl, uses reverse image search on memes identified so far in the crawl to identify other communities that have posted memes.
We iteratively crawl these new communities, again use reverse image search to identify additional communities that have also posted the images posted to those new communities, crawl those communities, and repeat until convergence.
This crawl produces a list of communities that we then scrape in the second stage.
Our high-frequency scrape captures every piece of content uploaded to any of those communities over one month. We again utilize image search to filter out memes that have been seen online prior to the sampling period, allowing us to only analyze memes that were created \textit{de novo} during our sampling period.
We follow prior work~\cite{zannettou2018origins} by using perceptual hashing and clustering to combine variations of the same meme template, merging variations of the same base image that feature different crops or text overlays.
\rev{Together, these methods allow us to identify the first posting of an image meme on the public web, which we identify as the \textit{originating} community, and track other communities that later posted the same meme image.}
We map each community onto a continuous measure of its position on the core-periphery axis via a network centrality measure~\cite{rochat2009closeness} calculated over a crawl of the entire web from the same period~\cite{commoncrawl}.

This process produced a dataset over one month of \num[group-separator={,}]{906481} meme images. With this dataset, we measure: (1)~Is the most-shared content \rev{originated} in the core or peripheral communities \rev{of the web's network structure}? 
(2)~Do core or peripheral communities produce more completely original content? 
This web-scale analysis allows us to analyze whether the cultural artifacts around us originate in small, peripheral communities \rev{of the web,} out of the public eye, and then diffuse to the core communities such as Reddit, Facebook, and Imgur, or whether modern cultural production has become centralized in the core.

Our results suggest that, while memes \camready{that} \rev{originated in more peripheral communities} \camready{will} diffuse farther individually, \camready{in aggregate} the vast majority of image meme diffusion events can be traced back to memes originating in the core of the network. For example, the top $10\%$ of communities by network centrality \rev{posted the original content that eventually goes on to represent $62\%$ of diffusion events online}. A zero-inflated negative binomial regression finds that core communities' memes are more likely to diffuse, but diffuse less far, when controlling for the size and activity level of the community. 

We then replicate our analysis recursively within a single influential community, analyzing the diffusions between core and peripheral subcommunities (subreddits) on Reddit. \rev{We find again that core subcommunities originate the memes that receive the vast majority of diffusions within the community}, but the mechanism is reversed: the core subcommunities' memes are less likely to diffuse individually, but diffuse farther when they succeed. This reversal of the web-scale result reinforces the importance of \camready{looking beyond} traditional within-community analyses in the literature \camready{and engaging in} cross-domain or web-scale replications.
We observe that all communities, including those in the core, \rev{can trace a plurality of their content back to other communities in the core.} \rev{Finally, \camready{we perform a} sensitivity analysis that confirms that our result is robust to even large amounts of unobserved memes diffusing from private or unobserved communities.} 

\rev{Web-scale analyses are rare in the literature, so it is important to articulate their strengths and limitations at the outset. In exchange for the benefits of offering an ecosystem-level view that generalizes beyond single platforms, web-scale analyses come with attendant methodological limitations because data must be captured and compared across as much of the web as possible.} 
\rev{Principally, our analysis cannot fully estimate the impact that private (unscrapeable) communities play, or rule out the possibility that some peripheral communities could be missed in the crawl. So, in this paper, we perform a sensitivity analysis and confirm that the magnitude of our result is substantially robust, so that it would still hold even if many extremely peripheral communities were missed. Second, }
image memes cannot capture all forms of culture: prior sociological work investigates other forms of cultural artifacts including music, literature, sports, and technology~\cite{lopes2002rise,hall2003sociology,ryan1950acceptance,kaufman2005cross,rossman2012climbing}.
However, image memes provide one valuable lens onto questions of cultural diffusion because they are among the most popular types of cultural artifact online, and they are easily created and transmitted, and they readily transfer across many communities and networks.
\rev{Third, operationalizing network centrality via web link network structure is not completely coincident with cultural network structure, since hyperlinks do not always equal cultural influence. However, at a macro scale, the internet link structure does capture a substantial amount of cultural influence, in the same sense that offline studies of cultural networks track social connections.}

This outcome suggests that the design affordances of modern social computing systems have centralized the publishing and diffusion of image memes online \rev{into the most popular and heavily linked-to sites online}. This outcome is not unique to the internet: other media such as radio and television have likewise exhibited patterns of centralization over time due to commercialization pressures~\cite{fish2017technoliberalism}.
Disruptive cultural innovations may still come from the periphery \rev{of the internet network}, of course, but our evidence suggests that the large plurality of the memes that we see online are originated in a small number of core communities. Being in the core offers large audiences and rapid re-sharing to other communities. Our results raise questions about what challenges we face as designers when cultural capital and innovation becomes increasingly centralized. 

We begin by deriving our hypotheses from the behavioral sciences. We then describe our dataset collection, and our analysis. Our dataset is available at \camready{\url{https://osf.io/hr3e5/}}, available for other researchers to download and use. 

\section[related]{Memes, Culture and Diffusion}\label{sec:related work}
Raymond Williams perhaps sets the stakes most effectively, writing that ``culture is one of the two or three
most complicated words in the English language''~\cite{williams2014keywords}.
Sewell declared two ways to define culture: a ``theoretically defined category or aspect of social life that must
be abstracted out from the complex reality of the human existence'' and as a ``concrete and bounded world of beliefs and practices''~\cite{bynum1999beyond}.
He goes further to conceptualize culture along many dimensions, ranging from learned behaviors, meaning-making, a system of symbols and meanings, and creativity.
In this work, we focus on cultural diffusion in its context attached to symbols that are spread online via image memes; however, sociologists have studied diffusion of many other facets of culture, cultural practices~\cite{kaufman2005cross}, technological innovation~\cite{ryan1950acceptance}, and linguistic evolution~\cite{leskovec2009meme}.

Much sociological work has focused on creation in the arts, such as jazz~\cite{lopes2002rise}, with generalization to other social behaviors: John Hall argues that ``there is little if any difference between the arts and other socially organized activities''~\cite{hall2003sociology}. The input is usually shaped by limited resources such as funding, and the output is shaped by regulation and censorship~\cite{lena2012banding}. 
There is also substantial evidence that cultural production does not arise from an idealized artist, but instead also features a large supporting cast ranging from friends giving feedback to a formal
organizations that mass produce the cultural item such as record labels and book publishers~\cite{hall2003sociology, rossman2012climbing}.
Success may depend on multiple self-reinforcing rounds of exposure to a new cultural artifact~\cite{mills2012virality}.

One important lens on cultural diffusion is that of network structure. The earliest work in cultural diffusion focused on technological innovations such as the acceptance of hybrid corn seeds~\cite{ryan1950acceptance}.
Diffusion processes of online content are likened to the way epidemics spread, with adoptees (infected individuals) spreading the innovation to adjacent nodes in the network~\cite{leskovec2006patterns,rodriguez2011uncovering,bauckhage2011insights}.
One major topic of interest is the route of diffusion: people preferentially share content that matches their beliefs~\cite{adar2005tracking}, and models can predict diffusion routes~\cite{ienco2010meme}. A second topic is the temporality of diffusion: sharing is often bursty initially, before slowing~\cite{myers2014bursty} and possibly recurring months or years later~\cite{cheng2016cascades}.
Research has translated many of these insights to online communities, such as studying the spread of videos on YouTube~\rev{\cite{figueiredo2014does,borghol2012untold,xu2016networked,xu2017longitudinal}} and TikTok~\cite{anderson2020getting}.

Image memes offer a modern context for studying forms of cultural innovation and diffusion~\cite{vickery2014curious,ling2021dissecting,10.1145/3449155}. They rise and fall quickly, with content continuously being flushed out of sight by new arrivals~\cite{bernstein20114chan}. Researchers have studied whether image meme diffusion cascades can be predicted~\cite{cheng2014can}, the temporal dynamics by which image memes come back and their cascades recur~\cite{cheng2016cascades}, the design choices that are associated with increased diffusion~\cite{10.1145/3449155}, and where image memes arise from within a limited set of communities~\cite{zannettou2018origins}.
We extend this literature by broadening the analysis from a small set of communities to essentially the entire indexable web, allowing us to test the impact that smaller, peripheral communities vs. large centralized communities play in this spread of cultural artifacts.

In sum, the existing literature has studied the structure of cultural diffusion, but rarely its genesis, and not at web scale. This motivates our research question:
\begin{resques}
    Are the online communities that originate popular image memes positioned in the core or periphery of the web?
\end{resques}

For our analysis, we draw on the practice of studying cultural innovation from the perspective of the structure of the social communities that produce it, often defined as \textit{core} and \textit{periphery}.
The core is defined as the mainstream, and the periphery as the marginal communities~\cite{bynum1999beyond}.
One major result in the literature is that, offline, the success of cultural innovation depends on the position in the social network, greatest when occupying an intermediate, bridging position neither in the core nor the
periphery~\cite{cattani2008core,doi:10.1287/orsc.1110.0673}.
The innovation's success also depends on its level of typicality relative to existing
content: the more diverse, the more likely it is to diffuse~\cite{berger2018atypical}.
However, the level of legitimacy attached to the person adopting the innovation also impacts their success rate~\cite{rao2003institutional}.

Prior work on individual or small sets of communities online suggests that communities that are well-known but still outside the core \rev{of the internet network} (e.g., 4chan) carry influence on the core communities, while core communities are more effective at pushing out memes to other places~\cite{zannettou2018origins}. Likewise, those in intermediate positions between core and periphery \rev{in professional or interaction networks} are most likely to create durable cultural innovations~\cite{cattani2008core,doi:10.1287/orsc.1110.0673}. If core communities are mostly \camready{echoing} content originated in the intermediate and peripheral communities, they should be responsible for fewer diffusion events:
\begin{hyp}
    \label{hyp:decreasing}
    Image meme diffusion events are negatively associated with network centrality, controlling for community size and activity level.
\end{hyp}



Another way we might test the literature's claim of the mainstream's inability to innovate~\cite{cattani2008core} is in the volume of original content (OC) that are originated in the community. This prompts the following hypothesis:
\begin{hyp}
    \label{hyp:OC}
    The rate of posting original content (OC) is negatively associated with network centrality, controlling for community size and activity level.
\end{hyp}

Online communities' structure and function are well studied in the CSCW research community~\cite{kraut2012building}, and memetic content arises regularly in their study (e.g.,~\cite{ling2021dissecting}). Memes arise as identity markers used for impression management~\cite{10.1145/3359170}, as members use them to project in-group signals. However, memes also present challenges for non-sighted community members~\cite{10.1145/3308561.3353792}.
Memes are also used regularly as vehicles for narratives in both social movements~\cite{10.1145/3134920} and disinformation campaigns~\cite{10.1145/3359229}.
Communities cannot always attend to the large volume of content posted to them~\cite{gilbert2013widespread}.
Less is known about the role that online communities play in producing online culture. Our research advances this goal through an analysis of the communities producing the dominant cultural icons.

\section[data]{Data and Methods}\label{sec:data}

\begin{figure}[b]
    \centering
    \begin{subfigure}[t]{0.33\textwidth}
        \centering
        \includegraphics[height=1.3in]{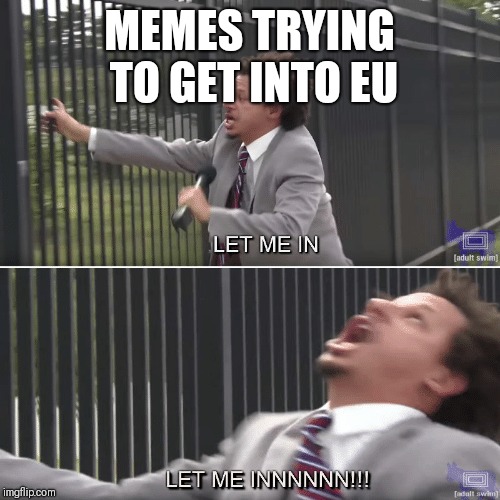}
        \label{fig:example_meme1}
    \end{subfigure}%
    \begin{subfigure}[t]{0.33\textwidth}
        \centering
        \includegraphics[height=1.3in]{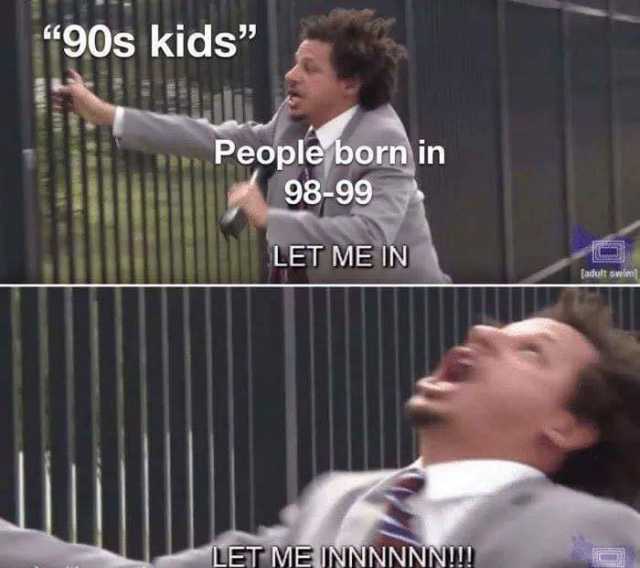}
    \end{subfigure}
    \begin{subfigure}[t]{0.33\textwidth}
        \centering
        \includegraphics[height=1.3in]{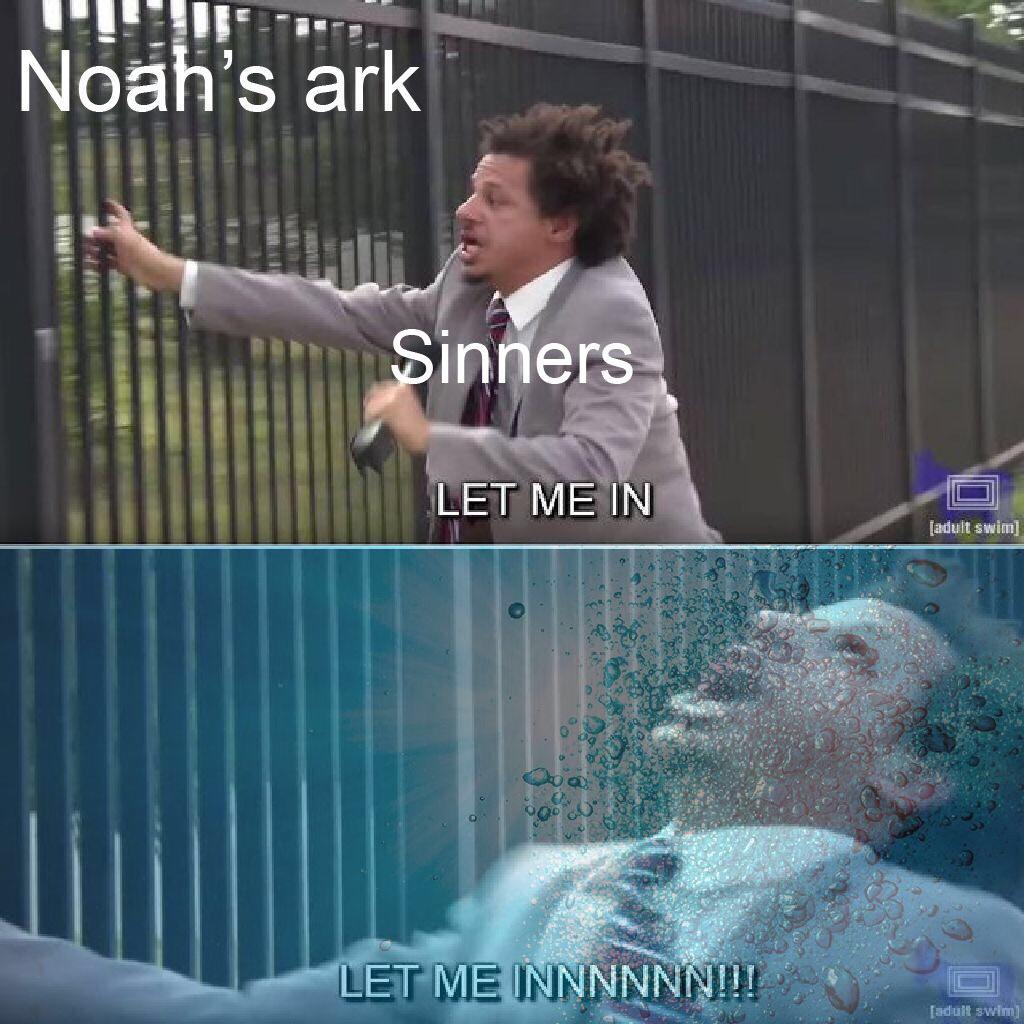}
    \end{subfigure}
    \caption{An example image meme. Our method automatically group memes with similar images and text, even accounting for differences in encoding, resolution, watermark, visual filter, or aspect ratio.}
    \label{fig:examplememe}
\end{figure}

In our study, we focus on image memes. Image memes (e.g., Figure~\ref{fig:examplememe}) are a dominant form of cultural artifact online today. They diffuse readily across many social media. Often memes contain both a base image and changeable text, or a base image where one part is replaced with another image, or are just a single image that gets shared widely.
We will focus on image memes that contain a base image with text overlaid.
We collapse instances of the same base image onto a single meme, allowing us to track that meme's origination and diffusion across the web even as the text overlay changes.

There are three main challenges in tracking image memes as they appear and are diffused across the web. First, we need to identify and track as complete a set of communities as possible, since the memes may be originally posted to small, peripheral communities that do not have many visitors. Second, we need to be able to identify memes as they move: image memes are transformed as they jump from community to community: they are downloaded and re-encoded; they change resolution; they acquire watermarks; they are re-cropped. Third, we need to restrict our analysis to only \textit{de novo} memes---memes which are new and have never been posted online before---filtering out memes that were posted prior to our study.

\subsection{Crawling communities}
To perform a web-scale analysis, we first need to identify a set of public communities on the web that we will track during the study period. Our crawl methodology identifies this set of communities by observing who is posting memes to the web, and then tracking its memes to identify other communities that have shared the same memes.
We make a simplifying assumption in treating a top-level domain as a community. Some communities, such as Facebook and Reddit, are very large and diverse. Others, as HiddenLol, are much smaller and focused. 

We identify our set of communities through a web crawl. This crawl tracks the sources of memes posted to known communities back to new communities that have not been observed yet, then adds those new communities to the set to track. We initialized our crawl with seed communities popular for their meme production: two subreddits (r\slash{}memes and r\slash{}funny), twelve Instagram accounts (e.g., memes, beigecardigan), 9gag's `fresh' feed, pinterest's `memes' pin, and accounts returned from Twitter's account search for `meme'. 
For each seed community, we wrote a custom scraper whose complexity ranged in complexity from a simple API client to a paginated HTML parser.
This scraper captured main content pages, or the posts within an account (e.g., on Pinterest).
We scraped a set of meme images from each community for review.

Then, for each meme image posted to our seed communities, we tracked that image backwards to other communities that had previously posted it. We achieved this by utilizing Google Cloud's Vision API~\cite{GoogleVision}, which provides a reverse image search to return other URLs where Google had indexed an identical or very similar image. We kept any domains that occurred at least ten times across our crawl. So, if several images in the sample had previously been posted on cscwmemes.com according to the cloud vision API, then cscwmemes.com was added to the set of communities, and its contents was then scraped and fed to the Cloud Vision API to help identify further new communities. \rev{By default, we included the top-level domain. However, in many cases, communities were better characterized by individual accounts or groups (e.g., subreddits on Reddit, accounts on Twitter), in which case we included those specific accounts or groups and attributed them to the top-level domain.}

Next, we filtered the communities to remove non-meme communities: not all potential communities are relevant sources of meme images. Our study is not focused on memes that arise as part of communities with other foci (e.g., a news community that occasionally posts a meme), so this filter allowed us to be certain that the images we were tracking were memes and not other images such as news photos. \rev{In addition, including non-majority meme communities would include non-meme images in our analyses, and which would confuse and dilute the analysis: e.g., capturing how photos from a news article spreads around the web. To maintain a coherent cultural lens, we focused on majority meme communities. The possible impacts of this decision are tested in our sensitivity analysis.} 

To achieve this, we utilized a crowdsourced labeling of these communities on Amazon Mechanical Turk to classify each one as posting a majority---over 50\%---meme content. Workers were paid \$0.30 per classification, to ensure at least minimum wage rates in our local jurisdiction. We presented workers with a definition of image memes that we were looking for, including examples, and asked three workers to independently classify whether each community we discovered featured a majority of image meme content. Workers were also asked to label if the site should be excluded for any of the following reasons: adult content, broken or private pages, non-English, news and advertising pages, online stores selling apparel with memes printed on them, or content hosted in private spaces such as Google Drive. These filters were included to ensure that our scrape would capture actual image memes, respected user privacy by only downloading public content, and would not download illegal underage pornography. Potential biases introduced by this process are considered in the Discussion section.

Once new communities had been vetted by this crowdsourced filter, we included them in our crawl and repeated the process, sampling their content and using the cloud vision API to identify additional communities or accounts within existing communities to include.
We repeated this recursive crawl until the set of communities converged. The full crawl period lasted several months.

Facebook, due to its API restrictions, was the one outlier in our crawl. Facebook disallows Google indexing of images in Facebook groups and pages. So, we began by filtering all public Facebook pages for those that contained the substring ``meme'' in their name. While Facebook pages could be included, Facebook groups are private and cannot be accessed except by members due to API restrictions, so we were unable to include them in our analysis. We discuss this limitation in Section~\ref{sec:limitations}; in brief, given the directionality of the outcome of our study, our conclusions would only be strengthened if Facebook groups were to be indexed as part of Facebook, because it would add additional memes originated to a community which is at the core of the network.

Our crawl ran from October 2018 through February 2019 and identified a total of 88 communities (domains). Communities included Reddit, Facebok, Dopl3r, Memedroid, Troll Street, Dump A Day, and many others. Most of these communities included one main feed; others comprised many meme accounts captured by our scraper, including \num[group-separator={,}]{1235} subreddits on Reddit, 929 Twitter accounts, 173 Instagram accounts, 51 Tumblr blogs, and \num[group-separator={,}]{1119} Facebook Pages.
We then manually added and whitelisted three communities that were featured centrally in prior work on meme culture---Reddit's r\slash{}the\_donald, as well as 4chan's \slash{}b\slash{} (random) and \slash{}pol\slash{} (politics) boards~\cite{bernstein20114chan,zannettou2018origins,keks}. While these communities are not primarily meme content, and were thus excluded by our filter, we manually whitelisted and included these communities in our dataset to maintain continuity with prior work; the results reported in this paper replicate when they are not included.
%
%
%
%
\subsection{Capturing diffusion events}
Having identified the set of meme-centric communities to track, the second part of our process collected a complete set of image memes with English-language text overlays posted to those communities over a period of a month. \rev{About three quarters of memes in our dataset contain text overlays. We remove memes without text, because textless memes often mutate via image editing software in ways that are more challenging to track. Including textless memes results in many false negatives where instances of the same meme are not recognized by the algorithms meant to group meme images together, because they look substantially different from each other.} \footnote{\rev{The Limitations section reflects on biases that this decision might introduce.}} We then merged instances of the same meme \textit{template} (shared base image) and filtered out memes that were posted elsewhere on the internet prior to our study period, ensuring that the same memes are grouped together in our dataset and each meme represents new original content created during the study period.

To achieve this, we performed a high-frequency data collection from each of the communities for a period of 30 days in March--April 2019. We elected for a 30 day sample because most meme lifecycles are much shorter than this~\cite{cheng2016cascades}.
For each image posted to any community in our list, we captured metadata such as date and time posted, title, description, and popularity via upvotes or retweets if available. It is important for our study of diffusion that we be able to accurately timestamp posts: so, we used techniques such as retrieving creation and modification timestamps via domain specific APIs, reading the timestamp from the HTML, using HTTP Headers to capture caching information, and if all the previous techniques failed, we fell back on the server's timestamp when the image was first crawled---we consider this to be accurate, as the crawling frequency was always within minutes.

Next, we must ensure that all memes in our dataset were \textit{de novo}, created and first posted during our study period and not before. To achieve this, we use the same Google Cloud Vision API~\cite{GoogleVision} reverse image search functionality from the crawling phase: when an image exists elsewhere in Google's search index, the Cloud Vision API returns a list of all prior locations of that image on the web. Any image memes that the Cloud Vision API returned results for prior to their first occurrence in our sample were excluded from our analysis, since the API returning results means that the image existed online before our study period.

\rev{We identified instances of the same meme following the same method as prior work~\cite{zannettou2018origins}. Specifically, we first run the perceptual hashing (pHash) algorithm~\cite{monga2006perceptual} on each image. pHash assigns a vector hash to each image, where images that appear similar based on a model of the human perceptual system will appear in similar parts of the pHash embedding. Then, following prior work~\cite{zannettou2018origins}, we use DBSCAN clustering~\cite{ester1996density} on the pHash results to group the pHashed images into clusters, where each cluster represents one meme. DBSCAN identifies sets of images whose pHash values are similar and groups them together. Each DBSCAN cluster collects one meme in its variations (e.g., Figure~\ref{fig:examplememe}). To achieve this, we replicate prior work's DBSCAN parameter settings~\cite{zannettou2018origins}.\footnote{These parameter settings from prior work resulted in reasonable clusters with the exception of one massive supercluster; we recursively divided this supercluster by requiring a tighter Hamming distance. A manual sampling of the top ten clusters confirmed that the resulting clusters represented different meme templates.}}


In summary: we first identify a set of communities online that post meme content, then scrape each image meme posted to any of these communities, then filter to the image memes created during our study period, then merged meme images that shared the same template. 
The outcome of this process is a set of memes, and a list of community timestamps for each meme when it was posted. The first timestamp for each meme represents the community that originated that meme: the meme existed nowhere else online before that post.

Our main dependent variable is the number of meme images in our sample that can be traced back to each meme. In other words, \rev{our construct of interest is \textit{diffusion volume}: the number of posts elsewhere that are attributable to a meme originated in the community.} So, for example, if a community posts three unique memes during the study period, the first meme might go on to get posted $0$ times in other communities \rev{(never diffused)}, the second meme might also get posted $0$ times, and the third meme might get posted $4$ times. We include reposts within each downstream community but not reposts within the originating community in this measure, because we are focused on external influence. We also measure a second dependent variable, the diffusion distance: the number of unique communities that each meme diffuses to. Because this measure is community-level, it excludes any reposts within the diffused communities. For example, in the earlier example, those three memes might have each diffused to $0$, $0$, and $2$ different communities. We consider diffusions between sub-communities (e.g., between two subreddits or Facebook Pages) to be within the same community.

These measures assume that the second chronological appearance of a meme captures a diffusion event from the first community to the second community.\footnote{It is possible that the same user posted to both communities; however, investigation of a sample of these diffusion events suggests that this is exceedingly rare.} Beyond this first diffusion event, we cannot know for sure which community is responsible: did the third event originate from the original community, or from the newer second community? However, 96\% of all diffusion events occur within one step of seed~\cite{goel2012structure}.
So, for our analysis, we focus only on attributing diffusions to the first community it appeared, rather than inducing an ordering on the following communities. 

We noted that some communities appeared to either scrape and auto-post, or to be intentionally collaborating in scraping or resharing each others' content. \rev{This behavior would artificially inflate counts, since the reshare is not a genuine cultural diffusion but an automated process.}
These situations were clear from their high volume and large proportion of cross-posting.
We removed any communities where both (1)~the volume from one community to another community was high, and (2)~a large proportion of memes on one community appeared on the second community. This criteria removed exactly two communities that were collaborating with each other---astrologymemes and loveforquotes. In addition, we noted that some communities in our crawl did not produce any new content in the period of the study, so we removed them from further analysis. \rev{Finally, we discarded any image memes whose text was not in English. Focusing on English served two purposes. First, it allowed us to focus on a coherent cultural context. Second, our research team and crowdsourced filter did not have the capacity to judge whether content is a meme across every written language, and financial constraints on the cloud vision API dictated that we could not consider all meme images in all languages.}  

\subsection{Mapping communities onto core and periphery}
Driven by the theory that inspired our work, we next mapped each community onto a measure of its network position, which captures how core (mainstream) or peripheral a community is \rev{in the world wide web link structure. The world wide web link structure, determined by the hyperlink graph of the web, captures many different sorts of influence in different links, including trust, citation, sourcing, related materials, and navigation. This paper concerns cultural (meme) diffusion; while the web link structure captures cultural influence (e.g., linking blogs, posts, and memes), the existence of other forms of link mean that the web link structure is more expansive than just cultural influence. This is a limitation of the wide lens of an internet-scale study. However, in this decision, we follow prior work studying core/periphery structures and artistic or innovation influence that also use networks that are related to but not identical with cultural influence, for example email threads in a consulting firm~\cite{doi:10.1287/orsc.1110.0673} and co-workers in films~\cite{cattani2008core}}.

We then apply a measure common in the literature, called \textit{network centrality}, to assign each community a continuous score of how core or peripheral it is on the web.
Communities in the core have high network centrality, communities in the periphery have low centrality, and intermediate communities are between \cite{easley2010networks}.
\rev{Prior work has used various centrality measures---e.g., dimensionality reduction via `coreness'~\cite{cattani2008core,doi:10.1287/orsc.1110.0673}---and we make no claim that our approach is the only reasonable one. However, we do require a centrality measure that can operate without being contaminated by link spam, link rings, and other issues that arise at web scale.}
A well-known measure of network centrality is PageRank~\cite{page1999pagerank}; however, it is susceptible to spam and manipulation. So, we adopted a common revision called harmonic centrality~\cite{rochat2009closeness}, endorsed and used by the Common Crawl foundation in their web-wide web crawl snapshot~\cite{commoncrawl}. We obtained harmonic centrality scores for each community from a full web crawl by Common Crawl completed during the same period as our data capture. In other words, we use a centrality measure derived from a full web crawl performed by Common Crawl to calculate how close each community is to the core or periphery of the web network. \rev{To test the robustness of our approach to other network centrality formulations, we also replicated our analysis using PageRank instead of harmonic centrality, finding the same results as reported in this paper.}

For interpretability, we standardized the harmonic centrality scores into z-scores, so that $0$ represents the average harmonic centrality, $1$ represents one standard deviation above the average, and so on.

For several of our analyses, we need to tag whether a post is original content (OC). OC are posts that are entirely \textit{de novo}, and have not been observed on the web previously. OC represents the new creative content that is uploaded onto each community. In our data, we annotated an image meme as OC if neither the cloud vision API nor DBSCAN identified a similar image meme posted previously on the web or in our dataset. Future instances of the same meme in our dataset are not considered OC.

We captured covariates to include in our analysis as well. These covariates included: (1)~the number of posts made on the community during the sample period, capturing the overall volume of the community; and (2)~community size, as measured by Alexa audience scores from \url{www.alexa.com}, a leading firm that estimates web traffic to domains across the web. These covariates were exponentially distributed, so we performed a logarithmic transformation on them to include in our regression models. Community size and number of posts exhibited some collinearity ($r=.65$); however, the model outcomes were stable with and without including community size, so we kept them both.

\section[results]{Results}\label{sec:results}
In total, we collected \num[group-separator={,}]{3308151} images, of which \num[group-separator={,}]{852453} were filtered out because they did not include text in them, \num[group-separator={,}]{525871} of them were removed because they didn't pass the safe search filter, and \num[group-separator={,}]{279872} were filtered out as non-English. After filtering, our dataset contained \num[group-separator={,}]{906481} images comprising \num[group-separator={,}]{683556} different image memes. Our dataset is available at \camready{\url{https://osf.io/hr3e5/}}, free for other researchers to download and use. The crawling code is available at \camready{\url{https://github.com/StanfordHCI/memescraper}}

\num[group-separator={,}]{13404} ($1\%$) of the memes diffused to at least one other community during our data collection period. The median diffusion distance amongst memes that diffused was three. The median time to first diffusion was 9.1 hours, suggesting that our thirty day scrape is of sufficient length to capture most diffusion events for a meme.

\rev{We organize our analysis into three main sections. The first two sections concern Hypothesis~\ref{hyp:decreasing}, that diffusion events are negatively associated with network centrality. Section~\ref{subsec:rq1} features the main analysis, across the whole web. To compare the results with a traditional single-community analysis, Section~\ref{subsec:rq2} then replicates the same analysis, but only on Reddit. Finally, Section~\ref{subsec:rq4} concerns Hypothesis~\ref{hyp:OC}, testing across the whole web the relationship between centrality and the probability of a meme being original content (OC).}

\subsection{Do image meme diffusions originate in core or peripheral communities \rev{on the web}?}\label{subsec:rq1}

\begin{figure}[tb]
    \centering
    \includegraphics[width=0.65\columnwidth]{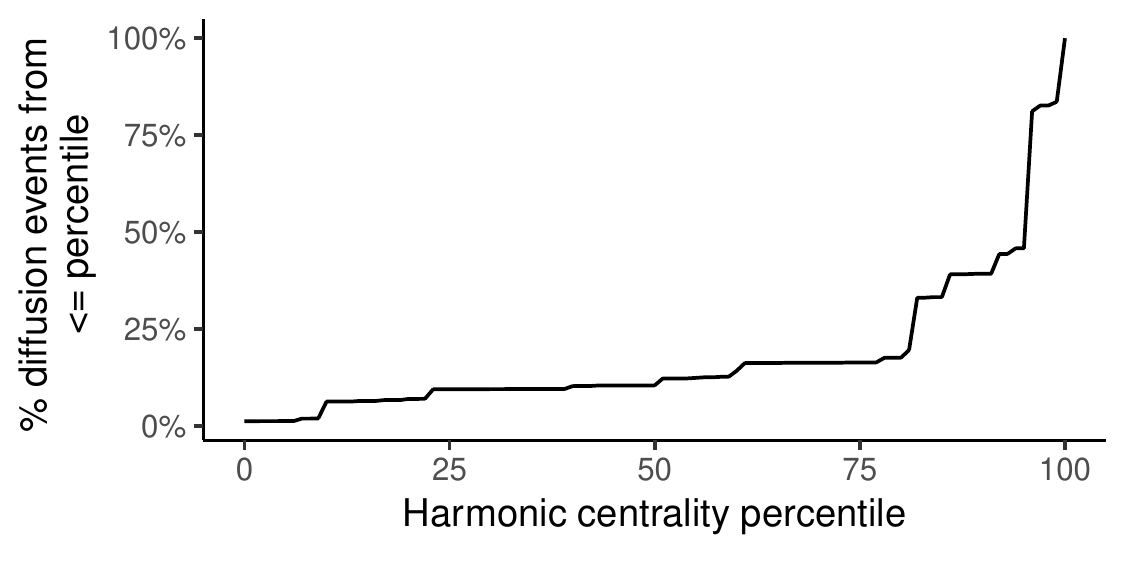}
    \caption{The majority of memes that diffused to other communities originated in  core communities rather than peripheral communities. For example, the top 10\% most core communities originate 62\% of diffusion events in our dataset.}
    \label{fig:diffusions_by_percent}
\end{figure}

\begin{figure}
    \centering
    \includegraphics[width=0.65\textwidth]{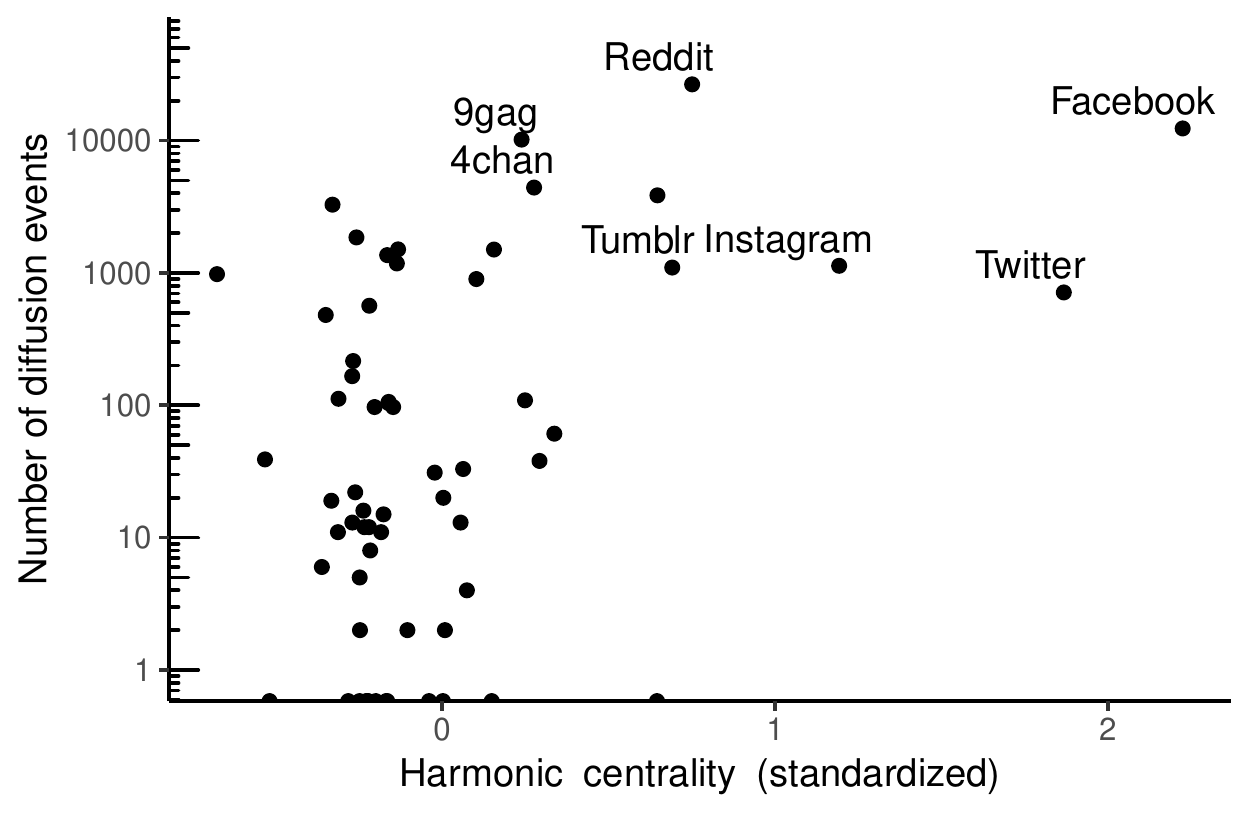}
    \caption{Communities with the highest harmonic centrality---including Reddit, Facebook, Twitter, and Instagram---were also among the most productive in terms of originating the most diffusion events.} 
    \label{fig:nummemes}
\end{figure}

\rev{In this section, we investigate our dataset to understand the relationship between centrality and diffusion events. We examine this question through several complementary methods. First, we examine descriptive statistics: what percent of diffusion events are attributable to memes originating in the core of the network? Second, we complement this descriptive result with a zero-inflated negative binomial regression, which helps us control for possible covariates. We then report the core and peripheral communities that originate the most (and fewest) diffusion events. Following this section, we will repeat this analysis within one large and influential community.}

The majority of image meme diffusion events on the web---62\%---originate from communities that are in the top 10\% as measured by harmonic centrality (Figure~\ref{fig:diffusions_by_percent}).
In contrast, about 20\% of meme diffusion events originate from the next 20\%, and about 18\% of meme diffusion events originate from the peripheral communities in the the lowest 70\%.

\begin{table}[tb]
    \centering
    \caption{Memes originating from more popular communities are more likely to diffuse, but diffuse less far.}
    \label{tab:diffusions_by_meme}
    \begin{tabular}{@{\extracolsep{5pt}}p{3.5cm}p{4cm}p{4cm}}
        \\[-1.2ex] & \multicolumn{2}{c}{\textbf{Diffusions by meme}} \\
        \hline \\[-1.8ex]
        & \makecell[l]{\textbf{Negative Binomial} \\ \footnotesize{\color{gray}{\rev{positive coefficient =}}}\\\footnotesize{\color{gray}{\rev{more diffusion events}}}} & \makecell[l]{\textbf{Zero-Inflated} \\ \footnotesize{\color{gray}{\rev{negative coefficient =}}}\\ \footnotesize{\color{gray}{\rev{more likely to diffuse}}}} \\
        \hline \\
        Harmonic Centrality (standardized) & $-$0.125$^{***}$ & $-$0.585$^{***}$\\
        & (0.027) & (0.019) \\
        Total Posts (log) & $-$0.105$^{***}$ & 0.134$^{***}$\\
        & (0.011) & (0.007)\\
        Active Members (log) & 0.028$^{***}$ & 0.052$^{***}$\\
        & (0.006) & (0.004)\\
        (Intercept) & 2.486$^{***}$ & 1.410$^{***}$ \\
        & (0.091) & (0.064) \\
        \hline \\[-1.8ex]
        Observations & 906,481 \\
        Log Likelihood & $-$105,898.500 \\
        \hline
        \hline \\[-1.8ex]
        \multicolumn{3}{c}{$^{*}$p$<$0.05; $^{**}$p$<$0.01; $^{***}$p$<$0.001} \\
    \end{tabular}
\end{table}

\rev{We performed a zero-inflated negative binomial regression, where each observation represents a meme and how many diffusion events it triggered. A zero-inflated negative binomial regression is a good fit for the data we are analyzing. Specifically, when modeling count data, it is generally ill-advised to use a linear regression, and more appropriate to use a method such as a Poisson regression, since a Poisson distribution better models counts of independent events. However, the Poisson distribution makes a strong assumption that mean equals variance---clearly inappropriate in a situation such as ours where some memes diffuse great distances---so it is common to utilize a negative binomial regression instead, which loosens this restriction. There is one final issue to address: the observation that 99\% of meme images in our data never diffused. The negative binomial regression does not correctly model data with overdispersed zeros such as this. So, we use a zero-inflated negative binomial regression, which corrects for this issue.}

\rev{Unlike a typical regression, which has a single set of coefficients, a zero-inflated negative binomial regression separates the regression into two sets of coefficients and estimates these two sets simultaneously. One set of coefficients, typically estimated as a logistic regression, models the probability that the data point will be a zero (never diffused). Larger (positive) coefficients for this zero-inflated factor indicate that increases in the predictor variable make a zero more likely. In other words, positive values mean that a meme is \textit{less} likely to diffuse. The other set of coefficients is the negative binomial component, which estimates the number of diffusion events for that meme. With the negative binomial, larger (positive) coefficients mean that the meme diffuses farther. So, a meme is most likely to diffuse, and to diffuse farthest, if zero-inflated coefficients are negative and negative binomial coefficients are positive.}

In the zero-inflated negative binomial regression, standardized harmonic centrality was associated with fewer zeroes ($z=-30.48, p<.001$), meaning that the more core the community, the more likely the meme was to diffuse at least once. Standardized harmonic centrality was negatively associated with how far the meme ultimately diffused ($z=-4.655, p<.001$, Table~\ref{tab:diffusions_by_meme}). In other words, the memes \rev{originated in} core communities diffuse less widely than memes \rev{originated in} intermediate and peripheral communities, but they are more likely to diffuse in the first place. Having more active users in the community \rev{was associated with a lower probability of diffusion} ($z=11.45, p<.001$), but \rev{with larger diffusion volume} ($z=4.56, p<.001$). More total posts were associated with content being more likely to diffuse ($z=17.07, p<.001$), and diffusing farther ($z=-9.64, p<.001$).

We investigated the communities in each category that were notable for their success (or lack thereof) in posting content that later diffused.
For communities in the core (here, above the 90th percentile in harmonic centrality), the community with the most diffused memes was Reddit, which originated \num[group-separator={,}]{26576} diffusion events to other communities. Facebook followed, with \num[group-separator={,}]{12346} diffusion events.
Sampling from intermediate communities (here, those in the 70th--89th percentile by harmonic centrality), those \rev{originating} the most diffusions were 9Gag and 4Chan; the \rev{one originating the fewest} was MemeGenerator. 
Within the peripheral communities (1st--69th percentile), the one \rev{originating} the most diffusions was HiddenLol; those \rev{originating the fewest diffusions} were JoyReactor, RuinMyWeek, Barnorama, MemeCollection, each with no diffusions over the month of data collection. 

We replicated this analysis with the diffusion distance rather than diffusion volume as the dependent variable. This second analysis counts the number of communities diffused to, rather than the number of raw meme images that were ultimately produced. Here, in the zero-inflated negative binomial regression, standardized harmonic centrality was still associated with fewer zeroes ($z=-29.62, p<.001$), but this time with slightly larger diffusion distance rather than shorter ($z=3.12, p<.01$).

These results provide mixed support for Hypothesis~\ref{hyp:decreasing} at the micro level of an individual piece of content, but not at the macro level of the ecosystem. Hypothesis~\ref{hyp:decreasing} predicts that more central sites will be less influential. With \textit{individual} pieces of content, higher harmonic centrality is associated with more diffusions, but shallower diffusions. However, the centralized sites make up for this in aggregate: even if individual pieces of content are diffuse less far, these communities push out so much content that they \rev{still originate} over 60\% of the diffusion events in our dataset. 
Peripheral communities may also have no \textit{a priori} interest in diffusing their content to the mainstream, explaining their fewer diffusions, but Section~\ref{subsec:rq4} will identify that even the peripheral communities draw much of their meme content from the core, so there is influence at least in one direction.

\subsection{Within communities, do memes originate from communities that occupy an intermediate network position \rev{on the web}?}\label{subsec:rq2}

Do these patterns hold within communities as well? In other words, is the web-scale analysis yielding different results than a traditional within-community analysis might? Are peripheral subcommunities within a core community acting more like peripheral or core communities? To answer these questions, we replicated our analysis on a single community---the social aggregator Reddit---now treating subreddits as individual communities. In Reddit, we observed which subreddit community originated a piece of content, and which other subreddits that content later appeared on.

\rev{This analysis follows a similar structure as the previous section. We analyze the overall percentage of diffusion events attributable to image memes originating in the core and periphery of Reddit. We then perform a zero-inflated negative binomial regression to control for additional covariates. We again then report communities that are especially notable for originating many (or few) diffusion events.}

\rev{However, to begin, we must construct the network graph amongst Reddit communities. With the web-scale analysis, we draw on the common practice of treating hyperlinks between webpages as directed edges. However, our preliminary experiments using links (e.g., crossposts) between Reddit communities demonstrated that Reddit has a much sparser sub-community link graph of explicit links. We sought a network structure that was valid to Reddit, using data available on Reddit.} 

\rev{We defined the directed edge weight as active commenter overlap between two communities. Consider a small source community $s$ and a large destination, default Reddit community $d$.  Intuitively, 75\% of the posters and commenters of the small community $s$ might also post or comment in the large default Reddit community $d$, indicating strong influence from $s$ to $d$, but only 0.1\% of the large default community $d$'s members would be active in the small community, indicating weak influence from $d$ to $s$. This is similar to the link structure of the web, where a small community might link to lots of content on Twitter, but only a small proportion of Twitter outgoing links are to the small community. Specifically, we defined the directed edge weight from subreddit $A$ to subreddit $B$ as active membership overlap, the number of users who commented in both subreddits during our data collection period divided by the total number of users who commented in subreddit $A$: $\frac{|A \cap B|}{|A|}$. Because the denominator $|A|$ changes depending on which direction the edge is pointing (e.g., $s \rightarrow d$ or $d \rightarrow s$), this is a directed graph. We calculated centrality again using standardized harmonic centrality.}

With the within-domain analysis, we were also afforded a more precise estimate of the number of active members of each community. So, for this analysis, our covariate for community size was measured as the number of unique commenters in the subreddit during our observation period.

This analysis uses the same dataset as the prior section, restricted to a single community and peering inside that community instead of collapsing the community to a single node as in the prior analysis. We again trained 
a zero-inflated negative binomial regression predicting the diffusion volume for each image meme.


\begin{figure}
    \centering
    \includegraphics[width=0.65\columnwidth]{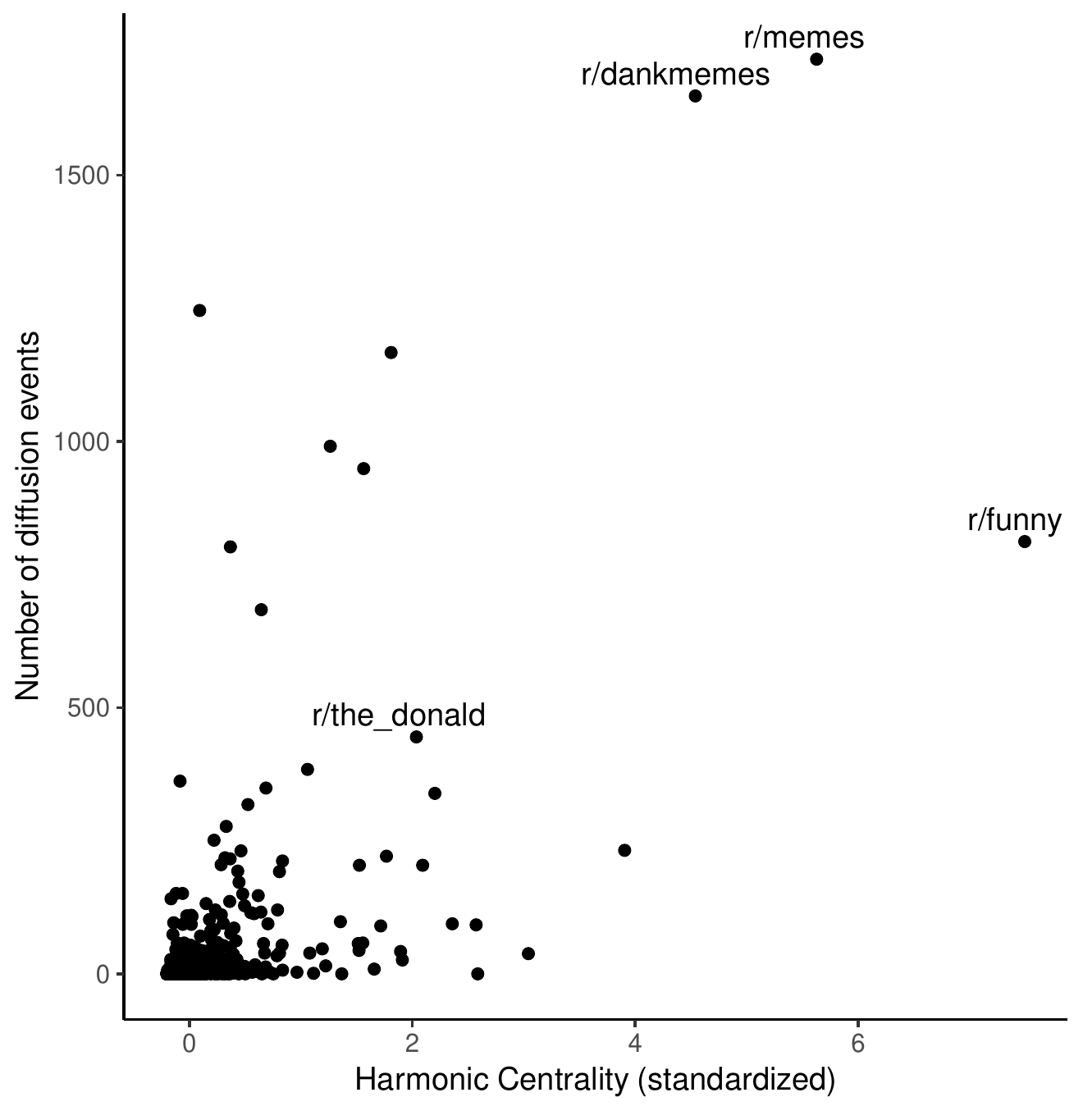}
    \caption{Similar to our cross-community results, the majority of memes that diffused between sub-communities on Reddit originated in the core of the Reddit network, rather than in intermediate or peripheral subcommunities.}
    \label{fig:nummemesreddit}
\end{figure}

The overall volume measurements replicated the across-community analysis. In Reddit, subreddits in the top 10\% by harmonic centrality were responsible for originating 69\% of diffusion events, the next 20\% by harmonic centrality originated 21\%, and the final 70\% in the periphery together originated 10\% (Figure~\ref{fig:nummemesreddit}).

In the zero-inflated negative binomial regression within this community (Table~\ref{tab:diffusions_by_meme_reddit}), the coefficients reversed compared to the cross-community regression. Standardized harmonic centrality was associated with more zeroes, meaning that the more central the subreddit, the lower the probability that the meme would diffuse to another subreddit. Standardized harmonic centrality was also associated with more diffusions, meaning that \rev{the memes originated in the more central subreddits} diffused farther, conditioned on them diffusing. So, the within-community analysis suggests potentially different mechanisms at play than the across-domain analysis, with the peripheral subcommunities within this core community \rev{originating memes that spread more} to the core subcommunities but not to as many other subcommunities compared to core subcommunities.

We again catalogued individual accounts or communities of note. On Reddit, in the core, the subreddits \rev{whose memes originated} the most diffusion events were r\slash{}{}dankmemes, r\slash{}{}historymemes and r\slash{}{}memes. Within intermediate subcommunities, r\slash{}{}deepfriedmemes \rev{originated} the most diffusion events, and r\slash{}{}comedycemetery \rev{originated} the fewest.
Within peripheral subcommunities, r\slash{}{}interestingasf*** and r\slash{}{}depression\_memes \rev{originated} the most diffusion events; r\slash{}{}explainitpeter and r\slash{}{}peopledyinginside \rev{originated} the fewest.

\begin{table}
    \centering
    \caption{Memes originating from more central subcommunities on Reddit are less likely to diffuse, but diffuse farther.}
    \label{tab:diffusions_by_meme_reddit}
    \begin{tabular}{@{\extracolsep{5pt}}p{3.5cm}p{4cm}p{4cm}}
        \\[-1.2ex] & \multicolumn{2}{c}{\textbf{Diffusions by meme on Reddit}} \\
        \hline \\[-1.8ex]
        & \makecell[l]{\textbf{Negative Binomial} \\ \footnotesize{\color{gray}{\rev{positive coefficient =}}}\\\footnotesize{\color{gray}{\rev{more diffusion events}}}} & \makecell[l]{\textbf{Zero-Inflated} \\ \footnotesize{\color{gray}{\rev{negative coefficient =}}}\\ \footnotesize{\color{gray}{\rev{more likely to diffuse}}}} \\
        \hline \\
        Harmonic \\ Centrality \\ (standardized) & 0.364$^{***}$ & 0.288$^{***}$\\
        & (0.081) & (0.053) \\
        log(OC posts) & $-$1.473$^{***}$ & 3.071$^{***}$\\
        & (0.358)  & (0.198)\\
        log(total posts) & 1.820$^{***}$ & $-$2.856$^{***}$\\
        & (0.345) & (0.190)\\
        log(users) & $-$0.339$^{**}$ & $-$0.313$^{***}$\\
        & (0.041) & (0.027)\\
        (Intercept) & 1.081$^{***}$ & 5.166$^{***}$ \\
        & (0.291) & (0.195) \\
        \hline \\[-1.8ex]
        Observations & 371,113 \\
        Log Likelihood & $-$37,593.440 \\
        \hline
        \hline \\[-1.8ex]
        \multicolumn{3}{c}{$^{*}$p$<$0.05; $^{**}$p$<$0.01; $^{***}$p$<$0.001} \\        
    \end{tabular}
\end{table}

Similar to cross-domain section, we replicated this analysis with the diffusion distance rather than diffusion volume as the dependent variable. The results replicated: 
in the zero-inflated negative binomial regression, standardized harmonic centrality was associated with more zeroes ($z=4.819, p<.001$) and with larger diffusion distance ($z=3.401, p<0.001$).

\begin{table}[]
    \centering
    \begin{tabular}{p{2cm}|p{2cm}p{2cm}p{2cm}p{2cm}}
        & \multicolumn{2}{l}{\makecell[l]{\textbf{Negative Binomial} \\ \footnotesize{\color{gray}{\rev{positive coefficient =}}}\\\footnotesize{\color{gray}{\rev{more diffusion events}}}}} & \multicolumn{2}{l}{\makecell[l]{\textbf{Zero-Inflated} \\ \footnotesize{\color{gray}{\rev{negative coefficient =}}}\\ \footnotesize{\color{gray}{\rev{more likely to diffuse}}}}} \\
        \textbf{Predictor} & Web-scale analysis & Single-community analysis & Web-scale analysis & Single-community analysis \\
        \hline \\
        Harmonic centrality & Negative & Positive & Negative & Positive
    \end{tabular}
    \caption{\rev{Comparing the web-scale and single community analyses highlights the differences visible only at the web-scale lens. While the web-scale analysis had the same macro-scale result---that central communities originated more diffusion events---the mechanisms through which this occurred differed. Across the web, more central communities were more likely to diffuse (fewer zeroes) but diffused less far; within a single community, more central communities were less likely to diffuse (more zeroes) but diffused farther.}}
    \label{tab:webvsreddit}
\end{table}

Ultimately, these results provide a partial replication of the cross-community data. \rev{We compare the two studies' results in Table~\ref{tab:webvsreddit}.} Like the cross-community study, the most core subcommunities \rev{originated} the vast majority of diffusion events. However, the mechanism through which this occurred was different: fewer memes that diffuse, but memes that diffuse farther, compared to the cross-community study finding that the core communities have more memes that diffuse but will diffuse less far. This result reinforces the importance of pairing typical within-community analyses from the literature with cross-domain replications. 

\subsection{Do core or peripheral communities have a higher rate of posting original content (OC)?}\label{subsec:rq4}

To understand whether more central communities were producing less novel content, \rev{Hypothesis \ref{hyp:OC}}, we used a logistic regression analysis with the dependent variable coding whether each meme is original content, and the community's standardized harmonic centrality as the independent variable. \rev{We also visualized where the non-original content is attributable to for each community.}

The logistic regression identified that the more core the community, the greater its log odds of posting OC ($p<.001$; Table~\ref{tab:oc}). The model suggests that the odds ratio of a post being original content is multiplied by roughly 1.3 for each standard deviation increase in harmonic centrality.
In other words, core communities are more likely to be posting fresh content, and Hypothesis~\ref{hyp:OC} is not supported.


\begin{table}[tb]
    \centering
    \caption{Core communities are more likely to post original content.}
    \label{tab:oc}
    \begin{tabular}{p{3.5cm}cc}
        \\[-1.8ex] & \multicolumn{2}{c}{Original Content (1=OC)} \\
        \hline \\[-1.8ex]
        Harmonic Centrality (standardized) & 0.272$^{***}$ & (0.007) \\
        Active Members (log) & 0.011$^{***}$ & (0.002) \\
        Total Posts (log) & 0.071$^{***}$ & (0.002) \\
        (Intercept) & 0.105$^{***}$ & (0.020) \\
        \hline \\[-1.8ex]
        Observations & 906,481 & \\
        Log Likelihood & $-$506,715.700 & \\
        Akaike Inf. Crit. & 1,013,439.0000 & \\
        \hline
        \hline \\[-1.8ex]
        \textit{Note:}  & \multicolumn{2}{r}{$^{*}$p$<$0.05; $^{**}$p$<$0.01; $^{***}$p$<$0.001} \\
    \end{tabular}
\end{table}

\begin{figure}[t]
    \centering
    \includegraphics[width=0.65\textwidth,keepaspectratio]{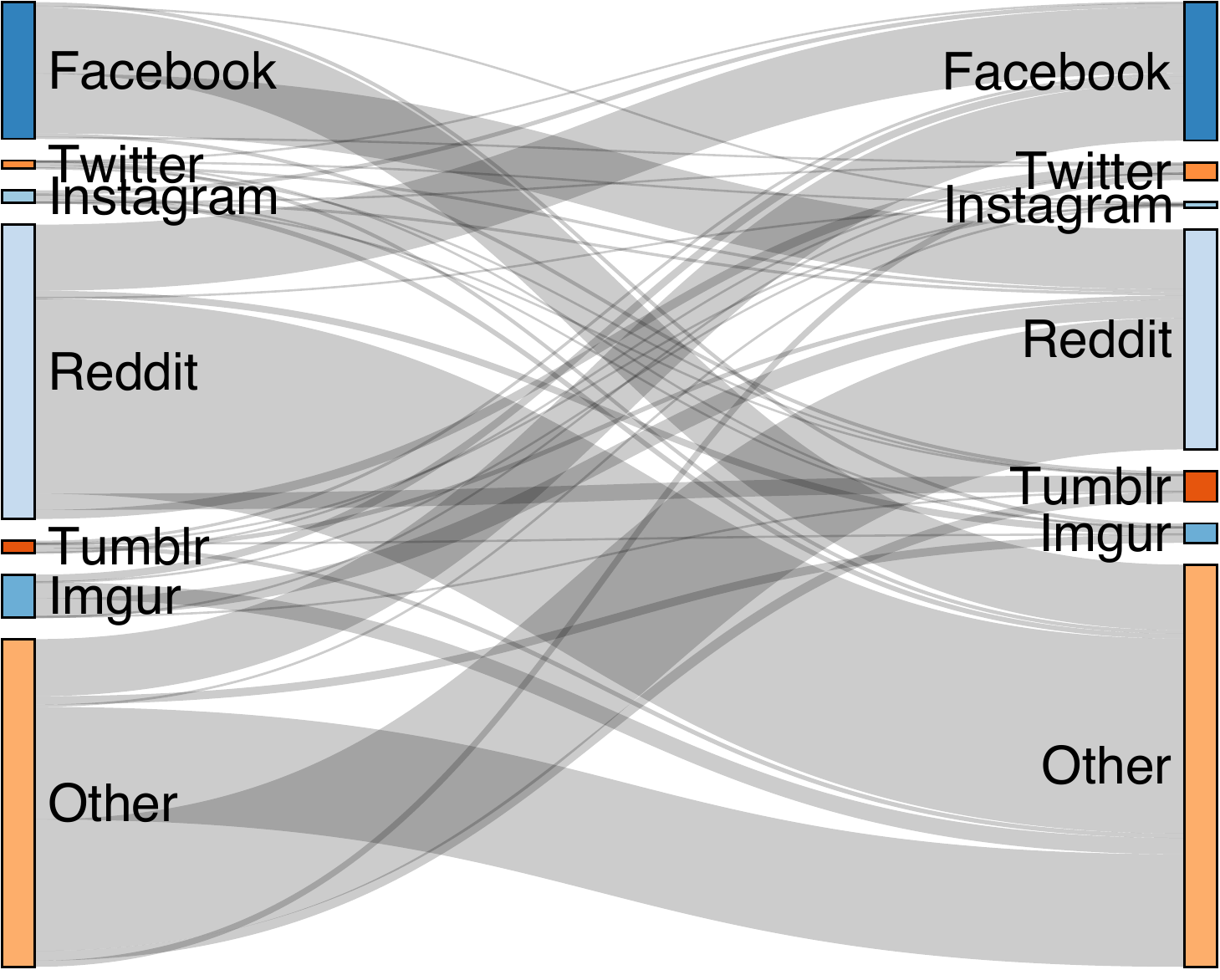}
    \caption{Generally, all communities draw much of their content from core communities. The left side of the Sankey diagram is where content originates; the flows to the right demonstrate the \rev{first observed diffusion location}. Content that \rev{originates} on Reddit or Facebook diffuses in all directions. Communities that are named in this visualization are those in the top 10\% by harmonic centrality.}
    \label{fig:sankey_domains}
\end{figure}

\rev{Which communities draw on content originated by which other communities? Figure~\ref{fig:sankey_domains} visualizes where memes originated in each community appeared next in our dataset. In general, all communities are posting a substantial amount of non-OC content that was originated in core communities, from Facebook and Reddit especially. About $\frac{2}{3}$ of the non-OC content posted in intermediate and peripheral communities was originated in just two core communities, Facebook and Reddit. However, about half of the non-OC content on Facebook and Reddit does originate in non-core communities.}

\rev{These results together suggest that core communities are more likely to repost content that originated elsewhere, but that the new content they do have is reposted in intermediate and peripheral communities. In addition, the core communities do post a non-trivial amount of content that was originated in intermediate and peripheral communities, but other content originated in other core communities is still the largest representation}.

\section[sensitivity]{\rev{Sensitivity analysis}}\label{sec:sensitivity}
\rev{Our approach may have missed some peripheral communities. This section asks: What if some of these communities originated influential content? Would this threaten the results?}

To test how missing peripheral communities might threaten our conclusions, we performed a sensitivity analysis. Specifically, we assumed a possible worst-case scenario where we did not find peripheral communities that were producing memes that originated a substantial number of diffusion events. To do this, we sampled all the communities from the top 10\% by harmonic centrality, excluding Facebook and Reddit as they were outliers even within the core communities---communities of this size were extremely unlikely to have been missed. We then labeled these new communities as peripheral by assigning them harmonic centrality equivalent to the smallest harmonic centrality existing in our dataset, and added these synthetic new communities to our dataset.

We continued doing this until the total volume of diffusions from the bottom 10th percentile was equal to the total volume of diffusions from the top 10th percentile. Achieving parity required adding 30 extra communities. Our original dataset comprised 88 communities, 56 of which produced content during our study, so by this analysis, we would have needed to miss a third of the meme-generating communities on the web through our crawl, and every single one of them would have needed to be both peripheral and equally productive as the most central communities such as Twitter and Pinterest. If this were true, these thirty communities would likely be as well known in the media and research literatures for \rev{originating} this content---so this alternative explanation seems unlikely. 

As another sensitivity analysis, we replicated this process again, this time trying to invalidate the zero-inflated negative binomial regression results for higher harmonic centrality having fewer zeroes. \rev{In other words, we again we sampled communities from the top 10\% by harmonic centrality, excluding Facebook and Reddit, assigning them harmonic centrality equivalent to the smallest harmonic centrality existing in our dataset, added these synthetic new communities to our dataset, and continued doing so in an attempt to make the harmonic centrality coefficient in the model nonsignificant.} After we added 100 communities, we ceased our attempt: the results were still significant.

\rev{In sum, by simulating increasingly worst-case scenarios for our crawling, we have strong evidence that it would be extremely unlikely for our results to be invalidated by missing influential peripheral communities. This result is due to the large effect sizes of the main results.}

\section[limitations]{Limitations}\label{sec:limitations}
\rev{A web-scale analysis provides a wide-angle lens that complements prior work's focus on individual communities, but it makes methodological tradeoffs to do so. Here, we reflect on these limitations.}

\rev{First, even though our crawl took place over months, we cannot make formal guarantees that we have captured every eligible community. This is a fundamental limit of wide-lens analyses such as web-scale crawls including ours; while the method brings benefits, it also brings its own shortcomings. In particular, if our crawl did not come across a reference link to the community, and if Google did not index the domain (e.g., due to a robots.txt exclusion), it may not have been found in our crawl. Likewise, for privacy reasons, we did not index communities and accounts that were private, for example WeChat, Facebook groups, and Discord channels. Some content would have been flagged by the cloud vision API because much of it does get archived, for example 4chan. However, other platforms such as WeChat do not. So, our analysis was limited to the first public sharing of a piece of content.} 

\rev{If the missing content is mainly from core communities (e.g., Facebook groups), the results here would only be strengthened, since the core would be originating an even higher percentage of the image meme diffusions. 
Substantial meme activity occurs within Facebook groups as well; however, we were not able to observe it. Extending these results to include Facebook groups would only be possible if one were able to gain access to internal Facebook data. However, Facebook already holds both high harmonic centrality and high influence in terms of the number of diffusion events it originates, so our results would only be strengthened. Our sensitivity analysis helps estimate the possible impact of any oversights in the opposite direction, of peripheral communities.}

\rev{However, if missing content were mainly from peripheral communities, it could undercut the results. Peripheral subcommunities within core communities are fine, because the indexing captured the large core communities; here we are mostly concerned with peripheral communities. This concern makes the sensitivity analysis central to the validity of our results. The sensitivity analysis found that our crawling procedure would have needed to miss tens of peripheral communities, each originating memes that diffused as much as those on Twitter and Pinterest. Such communities would have been known to the media, and thus to us. So, this threat is unlikely to undermine the results; if anything, the countervalent effect of missing core communities such as Facebook groups likely makes our results an underestimate of the core communities, not an overestimate.}

Second, we did not track communities that were not majority memes, with the exception of Reddit's r\slash{}the\_donald, 4chan \slash{}b\slash{}, and 4chan  \slash{}pol\slash{}, which were included for continuity with prior work~\cite{bernstein20114chan,zannettou2018origins,keks}. While we did account for memes posted first to a multipurpose community in our method---any memes that first appeared in a multipurpose community were flagged by the cloud vision API as having arisen earlier on the web when they appeared in our sample, and filtered out---multipurpose communities do play an important role in the ecosystem. However, because we wanted to focus our analysis on content that we could be confident was an image meme, we restricted our communities to those that Mechanical Turk workers voted as majority meme. \rev{Otherwise, the analysis would include images spreading for other purposes, e.g., images of political figures diffusing to other news sites reporting on the same story. This would weaken the construct validity of our measures. In addition, a practical constraint: each image \camready{sent} to the cloud vision API costs money, and including non-majority meme communities would require a much larger budget, since many of these communities are quite prolific but only a small proportion of their images would be of interest to the analysis.} Future work can extend our analysis by including additional communities and filtering out non-meme content. 

This limitation gives rise to a concern: non-majority meme sites behave differently than majority meme sites. For this reason, we manually whitelisted and included non-majority meme sites that featured prominently in previous analyses (e.g.,~\cite{zannettou2018origins}). These sites, such as Reddit's r\slash{}the\_donald, as well as 4chan's \slash{}b\slash{} (random) and \slash{}pol\slash{} (politics) boards, were included in our analyses. In contrast to prior work, they \rev{did not originate the content that was most widely diffused}: 4chan was ranked 4th in number of diffusion events in our per-domain analysis, and r\slash{}the\_donald was ranked 7th in number of diffusion events in our within-domain analysis. \rev{So, while non-meme communities play a non-trivial role in origination of cultural artifacts, they do not appear to be leading producers.}

\rev{Third, to map images onto templates, we excluded images that do not contain text. This resulted in dropping roughly 25\% of the images that we encountered. We made this decision because pHash replicates visual features that the human perceptual system is likely to group together, and textless memes are often mutated via image editing software in ways that the pHash algorithm cannot detect. In other words, including textless memes will introduce a large number of false negatives into the analysis, where instances of the same meme are treated as separate memes, and thus thought to not diffuse.}

\rev{However, removing textless memes may influence the results. For example, textless memes may spread in different social circles, or spread via different mechanisms, than memes with text. More broadly, the question might be posed as: how universal are our results, which focus on image memes with text, compared to other forms of cultural diffusion? For example, do our results \camready{generalize} to text phrases, hashtags, viral videos, fashion, and food? Following prior work~\camready{\cite{lopes2002rise,hall2003sociology,rossman2012climbing,cattani2008core,doi:10.1287/orsc.1110.0673}}, we can only focus on one such domain in our study. Relative to these other forms of cultural diffusion, we suspect that memes with text have more rapid lifecycles (they are faster to create and edit than even textless memes, which require skills such as Photoshop), and because of language barriers, may be more culturally bounded than textless memes.}


Fourth, we restricted this analysis to the English-speaking web, and the processes may differ in other cultural contexts. Practices of privacy, and of sharing, may not generalize. The cultural forces at play may differ in other contexts, and be more challenging to track if the sources cannot readily be scraped (e.g., WeChat in China).

\rev{Fifth, the network link structure of the web is not entirely coincident with social or cultural influence. As discussed in the Method section, we follow prior work in identifying core-periphery structure based on behavior (hyperlinking) that captures cultural influence, but also captures other elements. Our inspection of the harmonic centrality measure suggested that it does align with intuitions around core-periphery structure. However, there may be some room for differences. Our sensitivity analysis suggests that our results are robust to small (or even large) differences. However, it is important not to read too carefully into small differences in centrality score as being meaningful, as the score collapses together many forms of influence.}

Finally, all results here are correlational, \rev{as our research questions are primarily investigating a question of network structure and not of causation,} so we cannot draw any causal conclusions. For example, we cannot differentiate whether core communities cause greater diffusion, or whether the diffusion from these communities is what made them core in the first place. \rev{We also cannot identify the causal influence of the originating community, nor argue for agency of the originating community in any further spread.}

\section[discussion]{Discussion}\label{sec:discussion}

\subsection{\rev{Navigating methodological tradeoffs in web-scale analyses}}
\rev{The benefits of a web-scale study are it that it can consider an ecosystem-level view rather than focus narrowly on a single community, and that it can reveal patterns that a single community might not. However, performing such a study leads to methodological tradeoffs. For example, in our case, we made decisions to focus only on image memes with text, in English, posted by majority-meme communities. These decisions reflected different tradeoffs. Sometimes these tradeoffs were about construct validity: for example, including non-majority meme communities would mean that the crawl would eventually index news sites, blogs, and many other sites that would not qualify as ``communities'', and that most of the images being tracked would not be image memes---they might be news photos, logos, product photos, or a variety of other content. Sometimes these tradeoffs were about scale: every query to the image search API cost money; many centrality algorithms fail with link rings and other strategic behavior that arises at web scale. Sometimes these tradeoffs were about trading off false positives and false negatives: textless memes are mutated via image editing software in ways that pHash cannot detect, which would result in meme diffusions not being correctly attributed to the originating community. So, we chose to drop the 25\% of images that had no text. We made several of these tradeoffs of false positives vs. false negatives.}

\rev{No single method can capture the web in its breadth. Each community has its own norms, APIs, and format. Our general approach has been to identify a tight subset of material that can be compared across the ecosystem, as rigorously as possible. We deemed that admitting false positives into the dataset, e.g., non-meme images, was more dangerous to our method than restricting the dataset. However, we had to be careful that restricting the dataset was not overly biasing our results. In our case, for example, we felt there was not a strong reason to believe that the 25\% of non-text memes would behave categorically differently from the memes with text, and given that we could track the memes with text more successfully, we chose to focus on those.}

\rev{The other tradeoff is, of course, that no method can capture the web in its entirety. Even the most powerful search engines in the world respect robots.txt files and do not index every web site. Clearly, important content arises outside of their view. We again had to ask the question of whether this threatens our results. In our case, we followed a two-criteria principle: (1)~do we have a crawling algorithm that can identify the vast majority of content that exists?; (2)~can we quantify whether the missing content threatens our conclusions? Addressing (1) motivated our use of the cloud vision API and its search index, which could lead us to any other public domain we might have missed. Addressing (2) motivated the inclusion of a sensitivity analysis, deliberately trying to ``break'' the result.}

\rev{We hope that these reflections can inspire future work to continue to refine these principles.}

\subsection{\rev{Reflections on theory}}
We asked whether a web-scale analysis of image memes would replicate and extend prior offline work indicating that communities in intermediate positions \rev{originate} cultural production. A web-scale analysis attempts to characterize these dynamics more completely than is possible with traditional, single-community studies. \rev{Memes originated by the periphery were individually more likely to diffuse, but in aggregate, they were a relatively small percentage of memes that had diffusion events. This influence arises even if we control for activity level in terms of number of active participants. Peripheral and intermediate communities posted substantial content that was originated in core communities.} Ultimately, these results do not support Hypothesis~\ref{hyp:decreasing}. We also hypothesized that original content (OC) rates would be lower in the core than in the periphery. This hypothesis, Hypothesis~\ref{hyp:OC}, was also not supported.

Our research is positioned as complementing prior work, which is focused on individual communities, by zooming out to an ecosystem level. Our results suggest that this strategy can be a critical lens to understanding social computing, since we found that the mechanisms internally within a community did not completely match the mechanisms at web scale. While no study can completely chararacterize the entire web, and our study makes methodological tradeoffs to operate at web scale, our results suggest that the CSCW community can continue to push beyond studies of individual communities.

That core communities \rev{originate} so much of image meme culture online plays against popular and academic narratives. The most direct implication is that the cultural curve may have shifted online, with the ease of posting and sharing and the availability of large audiences all favoring mainstream and core communities moreso than is possible offline. Intermediate communities such as 4chan claim responsibility for a vast number of memes~\cite{bernstein20114chan,keks}; yet Reddit \rev{originates the content responsible for a far larger number of diffusion events. Core communities such as Reddit see large influxes of original content, and their content is also equally likely as peripheral communities to diffuse elsewhere.} The growth of content aggregators such as Reddit may suggest that peripheral communities are not frequented as much as they used to be earlier in web's history.

\subsection{\rev{Alternative explanations}}
Another possible explanation is that our dependent variable can be refined: image meme diffusions may be incremental cultural innovations, rather than deeper and more transformative innovations such as the \#metoo movement. Focusing on the communities responsible for memes at the 95th percentile of diffusion events might be a first step, but likely many of these innovations spawn multiple related memes. This remains future work.

While the popular narrative about marginal and counterculture communities is that they reject mainstream culture, peripheral communities appear to be sharing a substantial amount of content \rev{originated in} the core. It is possible that our network centrality measures are not able to distinguish marginal communities from simply unpopular ones. However, it is also possible that this sort of reverse diffusion happens offline as well, with less spotlight: peripheral communities do not necessarily reject all of mainstream culture. We do not analyze what these peripheral communities \textit{do} with this content---whether they use it earnestly themselves, mock it, or some combination---but this analysis confirms that the mainstream content influences their behavior.

\rev{Counter to prior work~\cite{zannettou2018origins}, we did not observe political communities such as Reddit's r\slash{}the\_donald and 4chan's \slash{}pol\slash{} to be originating memes that were responsible for an especially large number of diffusion events. Our data collection period was quite active in political news, too, covering Robert Mueller's delivery of his report on Russian meddling in the 2016 U.S. election and Robert Barr's announcement of his conclusion of no collusion in President Trump's campaign. This, of course, does not dispute that political communities have influence: however, it may help us place their broader place in the ecosystem.}

The social web has also begun to blur the meaning of core and periphery, as core communities such as Reddit play host to peripheral sub-communities such as r\slash{}interestingasf***. When an outsider hears that a meme originated at r\slash{}interestingasf***, for example, do they assume that it comes from the mainstream (Reddit), or from the periphery (r\slash{}interestingasf***)? Further within-community analysis would help examine these issues. 

Our research also makes clear a limitation of prior work, which is that we know little about the roles of peripheral subcommunities within core communities. Our study found that peripheral subcommunities behaved differently than peripheral communities, \rev{originating} more memes that diffuse, but those memes diffused less far. However, like the peripheral communities, these peripheral subcommunities \rev{did not originate} a large proportion of the overall image meme diffusions on the community writ large. The CSCW community can further investigate the role of these communities that are peripheral within popular platforms.

\subsection{\rev{Implications for design}}
A popular narrative is that much of internet culture comes from small peripheral communities. However, our work suggests that core communities have become not just the destination of memes, but a main origination point as well. Further, the core subcommunities are driving much of the diffusion within the core communities themselves---the core of the core appears to be extremely influential. We suggest that, recognizing this shift, core communities further support their members in creating and sharing cultural artifacts. How might they empower more people to engage in cultural artifact creation? For example, could these sites build their own simple image editing tools into the posts, or connect each meme image to its source so that it's easy for viewers to fork an image meme and create their own variant? Subcommunities such as Reddit's r\slash{}memeeconomy already expect this posting of the source template as a norm, but the site could support this behavior explicitly. 

However, there also exist design implications that contravene the former: how might we design for the periphery to increase their influence if desired? Are peripheral communities being increasingly cut out the cultural equation? How can we produce tools that allow smaller communities to find critical mass, produce cultural content, and when appropriate, share it? Or do the peripheral communities prefer to keep their symbols to themselves (e.g.,~\cite{peperacist})?

\section[conclusion]{Conclusion}\label{sec:conclusion}
Our results suggest that internet culture has bent the curve of cultural production toward the core. Toward this, we presented a web-scale analysis of image meme diffusion. Building on prior work in sociology, we hypothesized that cultural artifacts would arise from more peripheral communities in the network. We crawled meme communities across the web to identify a large number of communities and then captured a complete dataset of \num[group-separator={,}]{906481} instances of memes over 30 days. In contrast to our hypothesis, we found that \rev{memes originated in core communities \camready{originated the memes responsible for the majority of} diffusion events on the web}. 

\section{Acknowledgments}
The authors thank for their support: Stanford MediaX, the Stanford Cyber Initiative, the Google Cloud Credits program, and the Office of Naval Research (N00014-16-1-2894). Toyota Research Institute (``TRI'')  provided funds to assist the authors with their research but this article solely reflects the opinions and conclusions of its authors and not TRI or any other Toyota entity.

\received{January 2021}
\received[revised]{July 2021}
\received[accepted]{November 2021}

\bibliographystyle{ACM-Reference-Format}
\bibliography{references}

\end{document}